\documentstyle[12pt,/home/bijnens/tex/a4wide,epsf,rotate]{article}
\newcommand{\rcite}[1]{{\cite{#1}}}
\newcommand{\rref}[1]{{(\ref{#1})}}
\newcommand{\tref}[1]{{\ref{#1}}}
\newcommand{\rlabel}[1]{{\label{#1}}}
\newcommand{\rbibitem}[1]{\bibitem{#1}}

\newcommand{\be}{\begin{equation}}
\newcommand{\ee}{\end{equation}}
\newcommand{\ba}{\begin{eqnarray}}
\newcommand{\ea}{\end{eqnarray}}
\newcommand{\dis}{\displaystyle}
\newcommand{\mathrm}[1]{\mbox{\rm #1}}
\newcommand{\ovpi}{{\overline{\Pi}}}

\begin{document}
\begin{flushright}
NORDITA - 94/27 N,P
\end{flushright}
\begin{center}
{\Large\bf INTRODUCTION TO EXTENDED\\[0.3cm]
NAMBU-JONA-LASINIO MODELS\footnote{
Contribution to the Second DAPHNE Physics Handbook, eds. L. Maiani,
G. Pancheri and N. Paver.}}\\[2cm]
{\bf Johan Bijnens}\\[0.5cm]
NORDITA\\
Blegdamsvej 17,\\
DK-2100 Copenhagen \O, Denmark\\[1cm]
\begin{abstract}
A short introduction to quark models with four fermion terms and their
predictions for the parameters of low-energy effective Lagrangians is
given. Special attention is paid to predictions that are as general as
possible.
\end{abstract}
\end{center}
\section{Introduction}
\rlabel{intro}
The original Nambu-Jona-Lasinio model\rcite{NJL}
was an extension of the BCS-theory
of superconductivity to the domain of spontaneous symmetry breaking
in the strong interaction. The original model was phrased in terms of
nucleons but has been rephrased in terms of quarks in the seventies
by the work of Kleinert and others. In the mid-1980's it was revived once more
as a phenomenological model\rcite{ENJL}.
Recent extensive reviews can be found in
\rcite{review}.

The description given here will more or less present the model
from the point of view of the work that I have been involved in
\rcite{BBR,BRZ,BP1,BP2}.
This method tries to get as much as possible out of the underlying structure
of this class of models before putting in actual values of the parameters
and choosing a specific regularization. 

The main aim of this general class
of models is an attempt to understand the low energy parameters of
the Lagrangians described in chapter 2 from a model that is
somewhat closer to QCD with a minimal
amount of extra free parameters. The framework presented here can be
easily extended to include more nonlocality. A more general
treatment of the nonlocality can be found in the Quark Resonance 
model\rcite{Petronzio} or via attempts to approximately solve the
Schwinger-Dyson equations in QCD (see ref. \rcite{Cahill}
for a review). It also in some sense
includes a lot of the popular models like the chiral quark model(CQM)
\rcite{GM} and the QCD-effective action approach\rcite{ERT}.

I will first give a short description of the model and a few arguments about
its connection with QCD. Then the occurrence of spontaneous symmetry breakdown
will be discussed. In the next part the low-energy expansion and a few of the
relations between low-energy parameters that follow in general from it are
given. Last I will discuss a little how (Vector) Meson Dominance(VMD) finds
a basis in this way of looking at the low energy hadronic world.

\section{The model and its connection with QCD}

The QCD Lagrangian is given by
\ba
\rlabel{QCD}
{\cal L}_{\rm QCD} &=& {\cal L}^0_{\rm QCD} -\frac{1}{4}G_{\mu\nu}
G^{\mu\nu} \, , \nonumber\\
{\cal L}^0_{\rm QCD} &=& \overline{q} \left\{i\gamma^\mu 
\left(\partial_\mu -i v_\mu -i a_\mu \gamma_5 - i G_\mu \right) - 
\left({\cal M} + s - i p \gamma_5 \right) \right\} q \, .
\ea
Here summation over colour degrees of freedom 
is understood and
we have used the following short-hand notations:
$\overline{q}\equiv\left( \overline{u},\overline{d},
\overline{s}\right)$; $G_\mu$ is the gluon field in the 
fundamental SU(N$_c$) (N$_c$=number
of colours) representation; 
$G_{\mu\nu}$ is the gluon field strength tensor in
the adjoint SU(N$_c$) representation; $v_\mu$, $a_\mu$, $s$ and
$p$ are external vector, axial-vector, scalar and pseudoscalar
field matrix sources; ${\cal M}$ is the quark-mass matrix. 

All indications are that in
the purely gluonic sector there is a mass-gap. Therefore there seems to be
a kind of cut-off mass in the gluon propagator (see the discussion in 
ref. \rcite{Lattice}). 
Alternatively one can think of integrating out the 
high-frequency (higher than $\Lambda_\chi$, a cut-off of the order 
of the spontaneous symmetry breaking scale) gluon
and quark degrees of freedom and then expand the resulting effective
action in terms of local fields. 
We then stop this expansion after the dimension
six terms. This leads to the following effective action
at leading order in the $1/N_c$ expansion 
\ba
\rlabel{ENJL}
{\cal L}_{\rm QCD} &\rightarrow& {\cal L}_{\rm QCD}^{\Lambda_\chi}
+ {\cal L}_{\rm NJL}^{\rm S,P} + {\cal L}_{\rm NJL}^{\rm V,A} +
{\cal O}\left(1/\Lambda_\chi^4\right),\nonumber\\ 
{\rm with}\hspace*{1.5cm} 
{\cal L}_{\rm NJL}^{\rm S,P}&=&
\frac{\dis 8\pi^2 G_S \left(\Lambda_\chi \right)}{\dis
N_c \Lambda_\chi^2} \, {\dis \sum_{i,j}} \left(\overline{q}^i_R
q^j_L\right) \left(\overline{q}^j_L q^i_R\right) \nonumber\\
{\rm and}\hspace*{1.5cm}
{\cal L}_{\rm NJL}^{\rm V,P}&=&
-\frac{\dis 8\pi^2 G_V\left(\Lambda_\chi\right)}{\dis
N_c \Lambda_\chi^2}\, {\dis \sum_{i,j}} \left[
\left(\overline{q}^i_L \gamma^\mu q^j_L\right)
\left(\overline{q}^j_L \gamma_\mu q^i_L\right) + \left( L \rightarrow
R \right) \right] \,. 
\ea
Here $i,j$ are flavour indices and $\Psi_{R,L} \equiv
(1/2) \left(1 \pm \gamma_5\right) \Psi$. 
The couplings $G_S$ and $G_V$ are
dimensionless and ${\cal O}(1)$ in the $1/N_c$ expansion and summation 
over colours between brackets is understood. 
The Lagrangian ${\cal L}^{\Lambda_\chi}_{\rm QCD}$ includes only 
low-frequency modes of quark and gluon fields. The remaining
gluon fields can be assumed to be fully
absorbed in the coefficients of the local quark field operators 
or alternatively also described
by vacuum expectation values of gluonic operators (see the discussions 
in refs. \rcite{BBR,BRZ}).

So at this level we have two different pictures of this model. One is where
we have integrated out all the gluonic degrees of freedom and then
expanded the resulting effective action 
in a set of {\bf local} operators keeping only the first nontrivial
terms in the expansion.
In addition to this we can make additional assumptions.
If we simply assume that these operators are produced by the short-range part
of the gluon propagator we obtain $G_S = 4 G_V = N_c\alpha_S/\pi$. 
The
two extra terms in \rref{ENJL} have however different anomalous dimensions
so at the strong interaction regime where these should be generated there is
no reason to believe this relation to be valid. In fact the best fit
is for $G_S \approx G_V$. We report however also the fit with the constraint
$G_S = 4 G_V$ included.

The other picture is the one where we only integrate out the short distance
part of the gluons and quarks. We then again expand the resulting effective
action in terms of low-energy gluons and quarks in terms of local
operators. Here we make the additional assumption that gluons only
exists as a perturbation on the quarks. The quarks feel only the interaction
with background gluons. This is worked out by only keeping the vacuum 
expectation values of gluonic operators and not including propagating
gluonic interchanges. Most fits are in fact best with the gluonic
expectation value equal to zero (see table \tref{table1}).

This model has the same symmetry structure as the QCD action 
at leading order in $1/N_c$ \rcite{tHooft}
(notice that the $U(1)_A$
problem is absent at this order \rcite{Witten}).
(For explicit symmetry properties under SU(3)$_L$ $\times$
SU(3)$_R$ of the fields in this model
see reference \rcite{BBR}.) The QCD anomaly can also be consistently
reproduced\rcite{BP1}.

There has been a recent suggestion that the four-quark operators could result
from QCD ultraviolet renormalon effects \rcite{zakharov}. There it is suggested
that $G_V$ should be very small. The picture there is in fact quite different
from the one usually pictured in this model. They argue that the vector sector
follows more or less from the standard QCD picture but that the scalar and
pseudoscalar sector is strongly perturbed by ultraviolet renormalons. These
are local effects and can thus be described via a local four-quark operator
like the term proportional to $G_S$. Therefore I have included a fit to
the low-energy data with $G_V =0$. Other arguments for this model can be given
by looking at the local effects of purely gluonic operators. The most
prominent gluonic operator beyond $\langle G^2 \rangle$ are in fact
gluonic correlators of
$D^\alpha G_{\alpha\beta}$ which is related to four-quark operators. This
again leads to a NJL-like model at low-energies\rcite{BR}.
\section{Spontaneous Symmetry Breaking}

We can self-consistently solve the Schwinger-Dyson equation for
the fermion propagator in terms of the bare propagator and a 
one-loop diagram (see figure \tref{figure1}).
\begin{figure}[htb]
\unitlength 1cm
\begin{picture}(10,2)(-8,-0.5)
\thicklines
\put(-5,0){\vector(1,0){1}}
\put(-4,0){\line(1,0){1}}
\put(-2.5,-0.1){=}
\put(4,0){\line(1,0){1}}
\put(4,0.75){\circle{1.5}}
\put(4,0){\circle*{0.2}}
\thinlines
\put(3,0){\line(1,0){1}}
\put(2.5,-0.1){+}
\put(0,0){\vector(1,0){1}}
\put(1,0){\line(1,0){1}}

\end{picture}
\caption{The Schwinger Dyson equation for the propagator. A thin (thick) line
is the bare (full) fermion propagator.}
\rlabel{figure1}
\end{figure}
In the case where the current quark masses are set to 
zero this equation allows for
two solutions for $G_S > 1$, one with constituent quark mass
$M = 0$ and the other with  $M\ne 0$
and the model shows spontaneous chiral symmetry breaking.
In the presence of explicit chiral symmetry breaking only the second solution
is allowed. In the leading $1/N_c$ limit the solution of the Schwinger-Dyson
equation is a flavour diagonal matrix for the constituent quark masses with 
elements $M_{u,d,s}$.
The gap-equation now becomes
\ba
\rlabel{gap}
M_i &=& m_i - g_S \langle 0| :\overline{q}_i q_i : | 0\rangle \, ,
\\
\rlabel{VEV}
\langle 0 | : \overline{q}_i q_i : | 0 \rangle \equiv
\langle  \overline{q}_i q_i \rangle &=& - N_c 4 M_i 
\int \frac{{\rm d}^4p}{(2\pi)^4}
\frac{i}{p^2-M_i^2}
~~=~~ - \frac{\dis N_c}{\dis 16 \pi^2} 
4 M_i^3  \Gamma\left(-1, \epsilon_i\right) \, ,
\\ 
\rlabel{gs}
g_S &\equiv& \frac{4\pi^2 G_S}{N_c\Lambda_\chi^2}~.
\ea

Therefore, in this model the scalar quark-antiquark one-point function 
(quark condensate)  obtains a non-trivial nonzero value.
The dependence on the current quark-mass is somewhat obscured
 in eq. \rref{VEV}.
We use here a cut-off in proper time as the regulator.  The quantity $\epsilon_i$ appearing in 
\rref{VEV} is $M_i^2/\Lambda_\chi^2$. In figure \ref{Figvev} we have plotted
the dependence of $M_i$ on $G_S$ for various values of $m_i$
and $\Lambda_\chi = 1.160$ GeV.
\begin{figure}
\rotate[r]{\epsfysize=13.5cm\epsfxsize=8cm\epsfbox{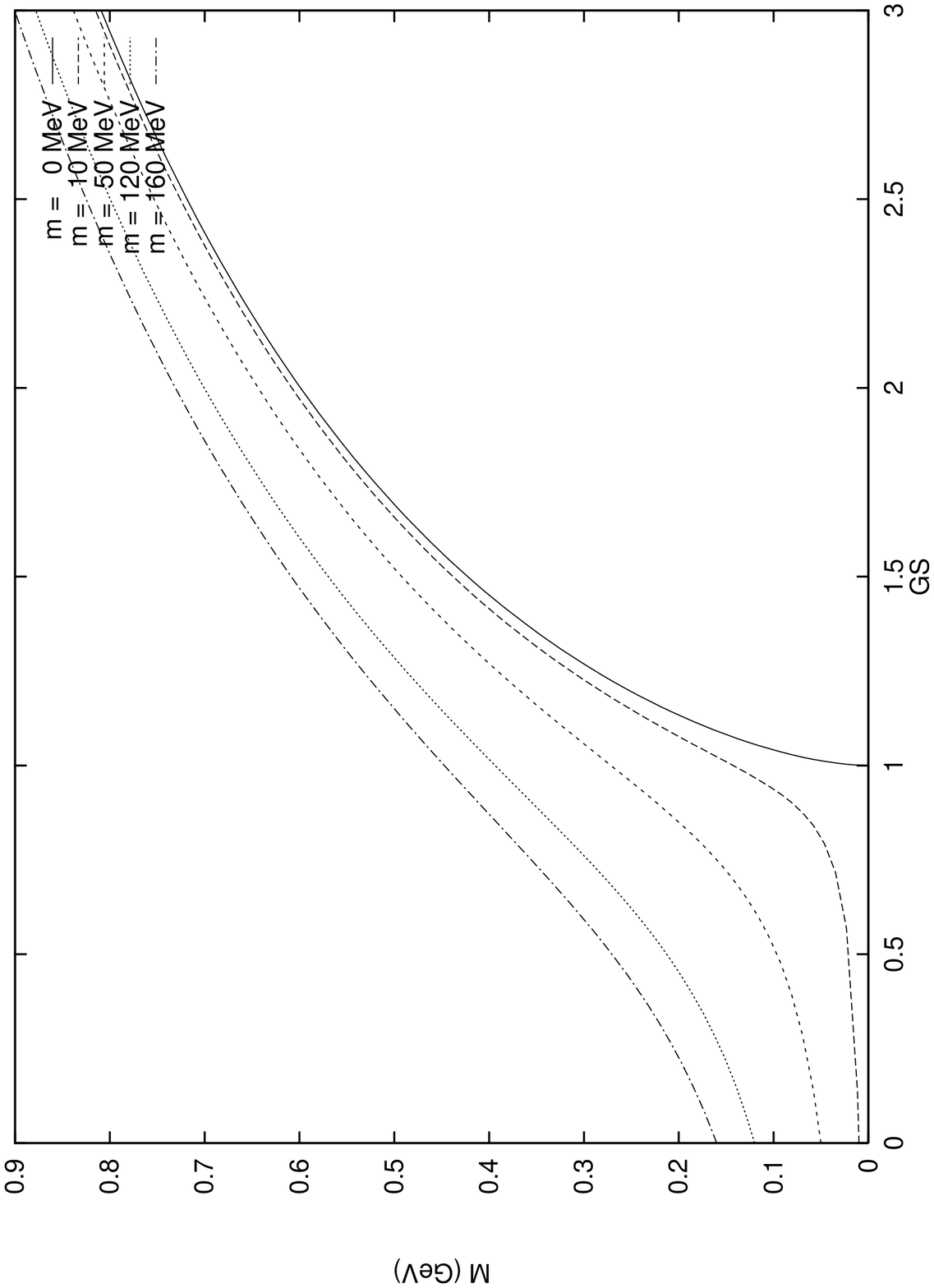}}
\caption{Plot of the dependence of the constituent quark mass $M_i$ as a
function of $G_S$ for several values of $m_i$}
\rlabel{Figvev}
\end{figure}
It can be seen that the value of $M_i$ for small $m_i$ converges smoothly
towards the value in the chiral limit for the spontaneously broken phase. 
This is an indication that an expansion in the quark masses 
as Chiral Perturbation Theory assumes for QCD is also valid in this model. 
However, it can also be seen that the validity of this expansion
breaks down quickly and for $m_i \simeq 200$ MeV we already
have $2 M_i \simeq \Lambda_\chi$.
We note that the ratio of vacuum expectation values
 for light quark flavours increases with increasing 
current quark mass 
at $p^2=0$ in this model and starts to saturate
for $m_i > 200$ MeV. 

In the 
mean-field approximation,
we can introduce the vacuum expectation value into the Lagrangian,
via an auxiliary field,
and then keep only the terms quadratic in
quark fields.
The ${\cal L}_{\rm NJL}^{\rm S,P,V,A}$ above
are then equivalent to a constituent chiral quark-mass term \rcite{ERT} of
the form $-M_Q (\bar{q}_L U^{\dagger} q_R + \bar{q}_R U q_L )$.

\section{The low-energy expansion}

This section is essentially a very abbreviated version of \rcite{BBR}. The 
Lagrangian \rref{ENJL} can be made into a form bilinear in quark fields by
introducing a set of auxiliary fields $L_\mu, R_\mu$ and $M$. The first
two transform respectively as a left (right) handed vector field (3 by 3
hermitian 
matrices) and the last one as $M \rightarrow g_R M g_L^{\dagger}$,
$M$ is a 3 by 3 complex matrix. Once the
Lagrangian has been brought into the bilinear form we can then assume
a vacuum expectation value for $M$ and make an expansion in its inverse.
This expansion can be done using the heat kernel expansion (see \rcite{Ball2}
for a review).
The vacuum expectation value is determined selfconsistently from this
expansion. This leads to the same gap equation (in the chiral limit) as
the one discussed in the previous section. Current quark mass dependence is
of course treated perturbatively with this method.

This procedure in fact generates kinetic terms for all the auxiliary fields.
So the model reproduces the low lying hadronic spectrum of pseudoscalars,
vectors, scalars and axial-vectors.
The combination $L_\mu+R_\mu$ becomes after a wave function renormalization
essentially the vector field. The different vector representations can be 
reached by making a redefinition
of the vector field here. 
Since the underlying model does not depend on the form of the vector field
chosen it is obvious that the resulting physics also does not depend on it.
(See \rcite{Ecker} for a discussion in the general case).
An added feature that appears here is that the
axial-vector auxiliary field, $R_\mu-L_\mu$ mixes with the
derivative of the pseudoscalar
field $U$. The $U$ field is obtained as the polar decomposition of $M$,
$M = \xi H\xi$. $H$ is hermitian and contains the vacuum expectation value
of $M$ and the scalar excitation part.
This mixing introduces a coupling of the pion field to the quarks which
is different from 1, the axial coupling of the (constituent) quark, $g_A$.
This already occurs at leading order in $1/N_c$. 
The
latter point is discussed exhaustively in \rcite{Peris}. This coupling
is smaller than 1 for $G_V\ne 0$. As an example of a successful prediction 
we have the value for $L_9$,
\be
\rlabel{L9}
L_9 = {N_c \over 16\pi^2}{1\over 6}\left[(1-g_A^2)\Gamma(0,x) +2
 g_A^2\Gamma(1,x)\right]\ .
\ee
$x= M_Q^2/\Lambda_\chi^2$ and $\Gamma(0,x)$ the
incomplete gamma-function.
In fact the main improvement is obtaining a better value for $L_5$ and $L_8$
where the inclusion of the vectors and scalar degrees leads to an improvement
over the QCD effective action model. The expressions are in fact rather simple.
\ba
L_5 &=& 
{N_c \over 16\pi^2}{1\over 4}g_A^3\left[\Gamma(0,x) -\Gamma(1,x)\right]\ ,
\nonumber\\
L_8 &=&{N_c \over 16\pi^2}{1\over 16}g_A^2\left[
\Gamma(0,x) -\frac{2}{3}\Gamma(1,x)\right]\ .
\ea
It is the presence of the extra $g_A$ factors and the cancellation between both 
terms that allow for phenomenologically good values for these latter two 
constants.

The main feature of this analysis is that even leaving the coefficients in
the heat kernel expansion completely free, leaves a number of interesting
relations. In this respect it should be mentioned that as a consequence
these relations survive in extensions of the model that do not
change the structure of the heat kernel terms. In particular, background
gluonic contributions do obey this criterion. These relations correspond
to a set of well-known phenomenological relations that were previously
derived using Meson dominance arguments and QCD short distance relations.
Some of them are
\ba
f_\pi^2  ~=~f_V^2 M_V^2 - f_A^2 M_A^2&~~~ &\mathrm{First Weinberg sum rule}
\nonumber\\
L_9 ~=~ \frac{1}{2}f_V g_V& ~~~&\mathrm{VMD of $\pi$ form factor}
\nonumber\\
L_{10} (2H_1) ~=~ -\frac{1}{4}f_V^2 +(-)\frac{1}{4}f_A^2&~~~&\mathrm{VMD of 
VV and AA 2-point functions}
\nonumber\\
g_A ~=~ 1 - \frac{f_\pi^2}{f_V^2M_V^2}\ .
\ea
The full list of relations can be found in \rcite{BBR}.
This approach also works very well numerically, witness table \tref{table1}.
The input parameters there are $M_Q$, $x = M_Q^2/\Lambda_\chi^2$, $g_A$
and the combination $g = \pi\alpha_S\langle G^2\rangle/(6 N_c M_Q^4)$.
Fit 2 is the fit with only low-energy parameters as input and everything free.
The meaning of the other fits can be found in \rcite{BBR} while the
last columns are with the low-energy parameters and the constraints
$G_V = G_S/4$(fit 6) and the renormalon picture with $G_V=0$.
We have also enforced the gluon condensate to vanish in these two fits.
Notice that in both cases the cut-off has decreased significantly
compared to the full case where $G_V$ was left free.

In general this model seems to interpolate well between VMD type of predictions
and chiral quark model type of predictions. This can be seen in eq. \rref{L9},
the first term is the one coming from the vector exchange while the second
one is the ``chiral quark loop'' contribution. By changing $g_A$ one can
have either a full VMD or a full CQM picture.
A major improvement compared to previous attempts in quark models
is the correct value
for $L_5$ and $L_8$ that are obtained here.
\begin{table}[htb]
\caption{Experimental values and predictions of the ENJL model
for various low-energy parameters from the heat kernel expansion.
All dimensionful quantities are in MeV. The difference
between the predictions is for slightly different choices of parameters.
(*) means that there are in addition uncertainties due to higher order chiral
corrections. 
The meaning of
the different fits are
explained in \protect{\rcite{BBR}}, fit 2 only uses $f_\pi$ and the $L_i$
while fit 6 and 7 include constraints on $G_V$, see text.
The numerical error in \protect{\rcite{BBR}} for $H_2$ has been
corrected. All masses are determined from the low-energy expansion,
not the pole position of the 2-point functions.}
\label{table1}
\begin{center}
\begin{tabular}{|c|c|c|c|c|c|c|c|c|c|c|}
\hline
    & exp. & exp.  & fit   & fit 1 & fit 2 & fit 3 & fit 4&fit 5&fit 6&fit 7\\
    & value& error & error &       &       &       &       & & &      \\
\hline
\hline
$f_\pi$ & 86(${}^{\dagger}$) &  $-$  & 10     &  89  &  86  & 86    &  87&83&
86&86\\
\hline
$\sqrt[3]{-<\overline{q}q>}$&194(${}^{\#}$)&8(${}^{\#}$)&$-$&281& 260 &  255
& 178 & 254 &210&170\\
\hline
$ 10^3 \cdot L_2$ & 1.2  &  0.4 &0.5  & 1.7   &  1.6 & 1.6 & 1.6 &1.7&1.5&1.6\\
$ 10^3 \cdot L_3$ &$-3.6$&  1.3 &1.3  & $-4.2$&$-4.1$&
$-4.4$&$-5.3$&$-4.7$&$-3.1$&$-3.0$\\
$ 10^3 \cdot L_5$ &  1.4 &0.5   & 0.5 & 1.6   &  1.5 &  1.1 
   &1.7&1.6&2.1&1.9\\
$ 10^3 \cdot L_8$ &  0.9 & 0.3  & 0.5 & 0.8   &  0.8 &  0.7    &1.1&1.0
 &0.9&0.8\\
$ 10^3 \cdot L_9$ &  6.9 &  0.7 &0.7  & 7.1   &  6.7 &  6.6   &5.8&7.1
&5.7&5.2\\
$ 10^3 \cdot L_{10}$&$-5.5$& 0.7 & 0.7&$-5.9$ &$-5.5$&$-5.8$&
$-5.1$&$-6.6$&$-3.9$&$-2.6$\\
$ 10^3 \cdot H_1$ & $-$  &$-$&$-$  & $-4.7$ &$-4.4$&$-4.0$&$-2.4$&
$-4.6$&$-3.7$&$-2.6$\\
$ 10^3 \cdot H_2$ & $-$  &$-$&$-$  & $ 1.4$ &$ 1.2$&$ 1.2$  &
 $ 1.0$&$ 2.3$&$-0.2$&0.8\\
\hline
$ M_V $&768.3 &0.5& 100 &    811  &  830 & 831     &  $-$ & 802&1260&$-$\\
$ M_A $& 1260&30&300 &     1331  & 1376 & 1609    &  $-$  & 1610&2010&$-$\\
$ f_V $& 0.20&(*)& 0.02&   0.18  & 0.17 & 0.17    &  $-$  & 0.18&0.15&$-$\\
$ g_V $&0.090&(*)& 0.009&    0.081  & 0.079&  0.079  & $-$ & 0.080&0.076&$-$\\
$ f_A $&0.097&0.022(*)& 0.022&   0.083  & 0.080& 0.068   & $-$&0.072&0.084&$-$\\
\hline
$ M_S $&983.3&2.6&200&            617  & 620  &  709    & 989 & 657&643&760\\
$ c_m $& $-$&$-$& $-$ &           20  &  18  &  20     & 24  & 25&16&6\\
$ c_d $&  34 &(*)&  10&          21  &  21  &  18     & 23   & 19&26&27\\
\hline
\hline
$x$  & & & &              0.052  & 0.063 &  0.057 &  0.089 & 0.035&0.1&0.2\\
$g_A$& & & &               0.61  &  0.62 &  0.62  &  1.0   & 0.66&0.79&1.0\\
$M_Q$& & & &                265  & 263   &  246   &  199   & 204&262&282 \\
$g$  & & & &                0.0  &  0.0  &  0.25  &  0.58  & 0.5&0.0&0.0 \\
\hline
\end{tabular}
\end{center}
\end{table}

\section{Beyond the low-energy expansion and Meson Dominance}
This section is a very short summary of references \rcite{BRZ} and \rcite{BP2}.
Similar work can be found in \rcite{Weise,Lutz,Meissner}.
For 2-point
functions the general graph is depicted in fig \tref{Fig2pt}.
\begin{figure}
\begin{center}
%
% F:\TEX\TEXDRAW\ENJL2.TEX
% Mainfile for graphic-inclusion (created by TeX-Draw, JP-91)
%
%
\thicklines
\setlength{\unitlength}{1mm}
\begin{picture}(140.00,35.00)(0.,15.)
\put(97.50,35.00){\oval(15.00,10.00)}
\put(103.00,33.50){$\bigotimes$}
\put(17.50,35.00){\oval(15.00,10.00)}
\put(25.00,35.00){\circle*{2.00}}
\put(32.50,35.00){\oval(15.00,10.00)}
\put(40.00,35.00){\circle*{2.00}}
\put(47.50,35.00){\oval(15.00,10.00)}
\put(55.00,35.00){\circle*{2.00}}
\put(62.50,35.00){\oval(15.00,10.00)}
\put(08.00,33.50){$\bigotimes$}
\put(68.00,33.50){$\bigotimes$}
\put(88.00,33.50){$\bigotimes$}
\put(38.50,19.00){(a)}
\put(95.50,19.00){(b)}
\put(14.50,40.00){\vector(1,0){3.00}}
\put(29.50,40.00){\vector(1,0){3.50}}
\put(44.00,40.00){\vector(1,0){5.00}}
\put(60.50,40.00){\vector(1,0){3.00}}
\put(95.50,40.00){\vector(1,0){5.00}}
\put(99.00,30.00){\vector(-1,0){3.00}}
\put(64.00,30.00){\vector(-1,0){3.00}}
\put(49.50,30.00){\vector(-1,0){3.50}}
\put(34.00,30.00){\vector(-4,1){2.00}}
\put(18.00,30.00){\vector(-1,0){2.50}}
\end{picture}
\caption{The graphs contributing to the two point-functions
in the large $N_c$ limit. 
a) The class of all strings of constituent quark loops.
The four-fermion vertices are either  
${\cal L}^{\rm S,P}_{\rm NJL}$ or  
${\cal L}^{\rm V,A}_{\rm NJL}$ in eq. \protect{\rref{ENJL}}.
The crosses at both ends are the insertion of the external sources. 
b) The one-loop case.}
\rlabel{Fig2pt}
\end{center}
\end{figure}
The sum of all diagrams is essentially a geometric series that can be
easily summed. The full two-point function
then becomes
\be
\Pi = \frac{\ovpi}{1-K \ovpi}~.
\ee
Here $\ovpi$ denotes the 2-point function at one-loop and $K$ the relevant
four-fermi coupling. When the 2-point functions mix a similar formula
exists except that $\Pi$ and $\ovpi$ become vectors and $K$ a matrix.
The rest
can be dealt with as a matrix inverse. 

Here we see how the resummation
has produced a pole and thus a bound state. In the previous section this
was dealt with by getting the kinetic term and then using the equations of 
motion for the auxiliary fields. The 2-point functions can usually be rewritten
in a meson-dominance form but with slowly varying parameters rather
than constants. E.g.(in the equal mass case) for the transverse vector case:
\ba
\Pi^{(1)}_V(Q^2) &=& \frac{2f_V^2(Q^2) M_V^2(Q^2)}{M_V^2(Q^2)-Q^2}\nonumber\\
2f_V^2(Q^2)M_V^2(Q^2) &=& N_c\Lambda_\chi^2 / (8\pi^2G_V)\nonumber\\
2f_V^2(Q^2) &=&\frac{8N_c}{16\pi^2}\int_0^1dx~x(1-x)\Gamma(0,x_Q)
\ea
with $x_Q = (M_Q^2 + x(1-x)Q^2)/\Lambda_\chi^2$,
and $Q^2 = -q^2$.

In fact some of the relations referred to in the previous section remain true
even after resummation to all orders. This is because the symmetries impose
certain identities on the one-loop functions and these have still some
consequences after resummation. Examples are the first and second Weinberg
sum rules, the famous $M_S = 2 M_Q$ relation and generalizations of these
away from the chiral limit \rcite{BRZ,BP2}.

A general argument for meson dominance appears here. An n-point function
consists of a set of graphs consisting of one-loop vertices joined by 2-point
functions. These two-point functions can now be rewritten in a form that
looks very much like meson dominance. Then as far as the ''vertices'' are
slowly varying we will find the VMD predictions within this model. In general
it is not so simple but still numerically results look very much like VMD,
see section 5 in \rcite{BP2}.

\section{Conclusions}
I have presented a short representation of the extended Nambu-Jona-Lasinio
model somewhat biased towards my own view of this set of models.
Further references can be found in the reviews cited. There is also work
on more vector meson phenomenology within the same approach \rcite{Ximo},
both for anomalous and non-anomalous decays.
The general conclusion is that within its limitations the ENJL-type models
do include a reasonable amount of the expected physics from QCD, its symmetries,
their spontaneous breakdown and even some of its short distance information (as
embodied in the Weinberg sum rules). It is also quantitatively successful.
Its major drawback is the lack of a confinement mechanism.
A general understanding of the interplay between meson dominance and the
chiral quark model is understood within this framework and a rather
good description of all low-energy parameters is obtained. 

\end{document}